\documentstyle[12pt]{article}
\begin{document}
\noindent

\begin{center}
{\Large {\bf  On Cosmological Implication of the Trace Anomaly }}\\
\vspace{2cm} ${\bf Y.~Bisabr}$\footnote{e-mail: ~y-bisabr@srttu.edu.} \\
\vspace{0.5cm} {\small {Department of Physics, Shahid Rajaee
University, Lavizan, Tehran 16788, Iran.}}\\
\end{center}
\vspace{1cm}
\begin{abstract}
We establish a connection between the trace anomaly and a thermal
radiation in the context of the standard cosmology.  This is done
by solving the covariant conservation equation of the stress
tensor associated with a conformally invariant quantum scalar
field. The solution corresponds to a thermal radiation with a
temperature which is given in terms of a cut-off time excluding
the spacetime regions very close to the initial singularity.  We
discuss the interrelation between this result and the result
obtained in a two-dimensional schwarzschild spacetime.
\end{abstract} \vspace{2cm}
\section{Introduction}
One of the important approaches to quantum gravity concerns a
framework in which matter is described by quantum field theory
while the gravitational field itself is regarded as a classical
object. In this framework, the so-called quantum field theory in
curved spacetime, the stress tensor associated with a quantum
field is not well-defined and contains singularities. Thus, some
renormalization prescriptions \cite{bd} are usually used to
obtain a meaningful expression for the stress tensor of a quantum
field. One of the most remarkable consequences of these
prescriptions is the so-called trace anomaly \cite{wald}. This
means that the trace of the quantum stress tensor of a conformal
invariant field obtains a nonzero expression while
the trace of the classical stress tensor vanishes identically.\\
It is shown \cite{cf} that in a two-dimensional schwarzschild
spacetime there is a close correspondence between the trace
anomaly and Hawking radiation \cite{haw}, namely the thermal
radiation associated with a black hole at null infinity. The
radiation has a temperature $T=(4\pi k_{B}R_{S})^{-1}$ where
$k_{B}$ is the Boltzmann's constant and $R_{S}$ is the
schwarzschild radius of a black hole.  It should be noted that the
length scale $R_{S}$ may be interpreted as a cut-off length
excluding the intrinsic singularity in the interior region of the
schwarzshcild solution.  In this sense, the temperature of the
Hawking radiation is given in terms of a cut-off length, $R_{S}$,
characteristic to the schwarzshcild spacetime.\\
The basic question we wish to address in the present work is that
whether there exists such a correspondence between the trace
anomaly and a thermal radiation in a cosmological context.  To
address this question we note that the trace anomaly is a
consequence of quantum behaviour of conformal invariant matter
systems, the characteristic of the matter systems at sufficiently
early times in the standard cosmology.  Following the result
obtained in the schwarzschild spacetime that the trace anomaly has
a long range effect at null infinity, one can also expect here
that it is possible to relate trace anomaly to properties of a
thermal radiation at late-time evolution of the universe.\\ Here
we intend to establish such a relation in two parts. In section
2, we first use a two-dimensional cosmological model to find the
most general solution of the covariant conservation equation of
the quantum stress tensor associated with a conformally invariant
scalar field. We shall show that the solution corresponds to an
equilibrium gas with a temperature $T\propto (k_{B}t_{c}^{-1})$
with $t_{c}$ being a cut-off time. This cut-off time is defined
to avoid the spacetime regions in which the semiclassical
investigations do not valid.  In section 3, we shall show that
contrary to the schwarzschild spacetime, the symmetries of the
standard cosmology allow us to solve the covariant conservation
equation in four dimensions.  We solve the conservation equations
for spatially flat Friedmann-Robertson-Walker spacetime.   At
late times, the solution corresponds to a thermal radiation with
a temperature of the same order of the temperature obtained in
the two-dimensional case.  In section 4,
we outline our results.\\
Throughout the following we shall
use units in which $\hbar=c=1$ and the signature is (-+++).\\

\section{The Model}
Let us begin with the results of renormalization of  stress
tensor $T_{\mu}^{\nu}$ of a quantum scalar field coupled with a
two-dimensional gravitational background

\begin{equation}
\nabla_{\nu } T_{\mu}^{ \nu}=0 \label{a1}\end{equation}

\begin{equation}
T_{\mu }^{\mu }=\frac{1}{24\pi}R\label{a2}\end{equation} where
$\nabla_{\nu }$ denotes a covariant differentiation and $R$ is the
curvature scalar.  The first equation is a covariant conservation
law and the second one indicates an anomalous trace emerging from
the
renormalization process.\\
We first intend to apply the conservation equation (\ref{a1}) to
a two-dimensional cosmological model described by the metric

\begin{equation}
ds^2=-dt^2+a^2(t)dx^2 \label{a3}\end{equation} where $a(t)$ is
the scale factor.  This is a two-dimensional analog of the
spatially flat Friedmann-Robertson-Walker (FRW) spacetime.  The
metric (\ref{a3}) can be written in a conformally flat form

\begin{equation}
ds^2=a^2(\tau)(-d\tau^2+dx^2) \label{a4}\end{equation} where

\begin{equation}
\tau=\int \frac{dt}{a(t)} \label{a5}\end{equation} is the
conformal time.  For the metric (\ref{a4}), the nonvanishing
Christoffel symbols are given by

\begin{equation}
\Gamma^{\tau}_{\tau\tau}=\Gamma^{\tau}_{xx}=\Gamma^{x}_{\tau
x}=\frac{1}{a}\frac{da}{d\tau} \label{a5'}\end{equation} In the
spacetime described by (\ref{a4}), all components of the stress
tensor can be only functions of time.  Using this fact, the
conservation equation (\ref{a1}) takes the form

\begin{equation}
\frac{d}{d\tau}T^{\tau}_{\tau}+\Gamma^{x}_{x\tau}T^{\tau}_{\tau}-\Gamma^{x}_{x\tau}T^{x}_{x}
=0\label{a6}\end{equation}

\begin{equation}
\frac{d}{d\tau}T^{\tau}_{x}+\Gamma^{\tau}_{\tau\tau}T^{\tau}_{x}-\Gamma^{\tau}_{xx}T^{x}_{\tau}
=0 \label{a7}\end{equation} For the off-diagonal elements of the
stress tensor we have $T^{\tau}_{x}=-T^{x}_{\tau}$. On the other
hand, one can write $T^{x}_{x}=T^{\mu}_{\mu}-T^{\tau}_{\tau}$.
These relations among different components of the stress tensor
together with (\ref{a5}) and (\ref{a5'}) allow us to write
(\ref{a6}) and (\ref{a7}) in the form

\begin{equation}
\frac{d}{dt}(a^2 T^{\tau}_{\tau})=a\frac{da}{dt}T^{\mu}_{\mu}
\label{a9}\end{equation}

\begin{equation}
\frac{d}{dt}(a^2 T^{\tau}_{x})=0 \label{a8}\end{equation} The
equation (\ref{a8}) gives immediately

\begin{equation}
T^{\tau}_{x}=\alpha ~a^{-2} \label{a10}\end{equation} with
$\alpha$ being an integration constant.  The solution of the
equation (\ref{a9}) is

\begin{equation}
T^{\tau}_{\tau}=a^{-2}(h+\beta) \label{a11}\end{equation} where

\begin{equation}
h=\int_{t_{c}}^{t} T^{\mu}_{\mu}(t')~\frac{da(t')}{dt'}~a(t')~dt'
\label{a12}\end{equation}

\begin{equation}
\beta=a^2 (t_{c}) ~T^{\tau}_{\tau}(t_{c})
\label{a13}\end{equation} and $t_{c}$ is an arbitrary time scale.
Given a time scale $t_{c}$, the function $h$ incorporates the
corresponding contribution of the trace $T^{\mu}_{\mu}$ in the
stress tensor $T_{\mu}^{\nu}$.  The introduction of the cut-off
time $t_{c}$ is a mandate in order to exclude in the definition
of $h$ the contribution of the trace very close to the early time
singularity.  In fact in that region an accurate description of
quantum gravity is needed and the semiclassical approach is no
longer valid.\\   In the context of the cosmology described by
(\ref{a4}), we require that $\alpha=0$.  This implies that the
stress tensor has vanishing off-diagonal elements.  In this case
we obtain

\begin{equation}
T_{\mu}^{\nu}=a^{-2}(h+\beta)\left(\matrix{1&0\cr 0&q-1\cr}\right)
\label{a14}\end{equation} where

\begin{equation}
q=\frac{T^{\mu}_{\mu}}{a^{-2}(h+\beta)} \label{a14'}\end{equation}
We are particularly interested in the late-time configuration of
the stress tensor.  It obviously  depends on the explicit form of
the scale factor.  Thus for studying $T_{\mu}^{\nu}$ at late
times, we first assume that the scale factor follows a power law
expansion

\begin{equation}
a=a_{0}(\frac{t}{t_{0}})^n \label{a15}\end{equation} with $t_{0}$
being the present age of the universe.  We then use the explicit
form of the trace anomaly for the metric (\ref{a4}) and in terms
of the scale factor (\ref{a15}) ( see the Appendix ) to write
(\ref{a12}) and (\ref{a14'}) in the form

\begin{equation}
h(t\rightarrow t_{0})=\frac{n^2~ a_{0}^2~
t_{0}^{-2}}{24\pi}\{1-l^{2(n-1)}\} \label{a16}\end{equation}

\begin{equation}
q(t\rightarrow
t_{0})=\frac{2(n-1)}{n}\{1+\frac{n-2}{n}~l^{2(n-1)}\}^{-1}
\label{a17}\end{equation} where $l=\frac{t_{c}}{t_{0}}$. Using
(\ref{a13}) and (\ref{a16}), the stress tensor (\ref{a14}) takes
the form
\begin{equation}
T_{\mu}^{\nu}(t\rightarrow
t_{0})=\frac{n^2~t_{0}^{-2}}{24\pi}\{1+\frac{n-2}{n}l^{2(n-1)}\}\left(\matrix{1&0\cr
0&q(t\rightarrow t_{0})-1\cr}\right) \label{a18}\end{equation} One
should note that the cut-off time $t_{c}$ is much smaller than
$t_{0}$ so that $l\ll 1$.  One therefore infers that $l^{2(n-1)}
\gg 1 $ for $n<1$. In this case, we may apply the approximation
that $q(t\rightarrow t_{0})\ll 1$ which reduces (\ref{a18}) to

\begin{equation}
T_{\mu}^{\nu}(t\rightarrow t_{0} )=\frac{n(2-n)~ l^{2n}~
t_{c}^{-2} }{24\pi}\left(\matrix{-1&0\cr 0&1\cr}\right)
\label{a18'}\end{equation} When one compares (\ref{a18'}) with the
stress tensor of an equilibrium gas, namely

\begin{equation}
\frac{\pi}{6}(k_{B}T)^2 \left(\matrix{-1&0\cr
0&1\cr}\right)\label{a19}\end{equation} one concludes that at
late times the stress tensor $T^{\nu}_{\mu}$ describes an
equilibrium gas with temperature
\begin{equation}
T= \frac{1}{2\pi}\sqrt{n(2-n)}~ l^n ~(k_{B}t_{c})^{-1}
\label{a20}\end{equation} It is interesting to compare
(\ref{a20}) with the result obtained in the schwarzschild
spacetime.  In that case, the temperature of the thermal
radiation is given in terms of $R_{S}^{-1}$. The schwarzschild
radius $R_{S}$ may be interpreted as a cut-off length that
disjoints the interior and the exterior of the schwarzschild
solution. In principle, this is very similar to the case of FRW
cosmology. In this case, the temperature of the equilibrium gas
is given in terms of $t_{c}^{-1}$.   Here the cut-off time
$t_{c}$ excludes the early stages of evolution of the universe in
which the semiclassical calculations can not be applied.
\section{The four-dimensional case}
In the standard cosmology, the universe is assumed to be isotropic
in all points of spacetime.  This is a larger symmetry with
respect to the schwarzshcild spacetime that allows us to
generalize our results obtained in the previous section to four
dimensions. In a four-dimensional spacetime, the analog of the
equation (\ref{a2}) is

\begin{equation}
T^{\mu}_{\mu}=-2v_{1}(x)\label{b2}\end{equation} where
\begin{equation}
v_{1}(x)=\frac{1}{720}(\Box
R-R_{\mu\nu}R^{\mu\nu}+R_{\mu\nu\gamma\delta}R^{\mu\nu\gamma\delta})
\label{b3}\end{equation} Here $\Box \equiv
g^{\mu\nu}\nabla_{\mu}\nabla_{\nu}$, $R_{\mu\nu}$ and $R$ are the
first and the second contraction of the Riemann curvature tensor
$R_{\mu\nu\gamma\delta}$, respectively.  We apply the
conservation equation (\ref{a1}) to the spatially flat FRW metric
described by
\begin{equation}
ds^2=-dt^2+a^{2}(t)(dx^2+dy^2+dz^2) \label{b4}\end{equation} or
in a conformally flat form
\begin{equation}
ds^2=a^{2}(\tau)(-d\tau^2+dx^2+dy^2+dz^2) \label{b5}\end{equation}
where the conformal time $\tau$ is given by (\ref{a5}).  The
nonzero Christoffel symbols of the metric (\ref{b5}) are
\begin{equation}
\Gamma^{\tau}_{\tau\tau}=\Gamma^{\tau}_{ii}=\Gamma^{i}_{\tau i}=
\frac{1}{a}\frac{da}{d\tau} \label{b6}\end{equation} where $i$ is
a collective index denoting three space components.\\
We intend to find the most general solution of (\ref{a1}) for the
metric (\ref{b5}).  To do this, one should note that in this case
all the components of the stress tensor $T_{\mu}^{\nu}$ are space
independent and can be only functions of time.  We use this fact
to write the time-component of the conservation equation
(\ref{a1}) as
\begin{equation}
\frac{d}{d\tau}T^{\tau}_{\tau}+3\Gamma^{i}_{i\tau}T^{\tau}_{\tau}
-3\Gamma^{i}_{i\tau}T^{i}_{i}=0 \label{b7}\end{equation} There
are also three space-components that all have the same form

\begin{equation}
\frac{d}{d\tau}T^{\tau}_{i}+\Gamma^{\tau}_{\tau\tau}T^{\tau}_{i}+3\Gamma^{i}_{i\tau}T^{\tau}_{i}
-\Gamma^{i}_{\tau i}T^{\tau}_{i}-\Gamma^{\tau}_{ii}T^{i}_{\tau}=0
\label{b7'}\end{equation} Note that there is no summation rule in
the relations (\ref{b6}), (\ref{b7}) and (\ref{b7'}).  In these
relations, one should simply replace the index $i$
with each of the space components, namely $x$, $y$ and $z$.\\
If we use (\ref{a5}) and (\ref{b6}),  the equations (\ref{b7}) and
(\ref{b7'}) take the form

\begin{equation}
\frac{d}{dt}(a^4 T^{\tau}_{\tau})=a^3 \frac{da}{dt}T^{\mu}_{\mu}
\label{b10}\end{equation}

\begin{equation}
\frac{d}{dt}(a^4 T^{\tau}_{i})=0 \label{b11}\end{equation} in
which we have also taken into account that
\begin{equation}
T^{i}_{i}=\frac{1}{3}(T^{\mu}_{\mu}-T^{\tau}_{\tau})\label{b10'}\end{equation}
and $T^{i}_{\tau}=-T^{\tau}_{i}$.  The equations (\ref{b11}) and
(\ref{b10}) yield
\begin{equation}
T^{\tau}_{i}=\delta ~a^{-4} \label{b13}\end{equation}
\begin{equation}
T^{\tau}_{\tau}=a^{-4}(H+\gamma) \label{b12}\end{equation} where

\begin{equation}
H=\int_{t_{c}}^{t}
T^{\mu}_{\mu}(t')~\frac{da(t')}{dt'}~a^3(t')~dt'
\label{b14}\end{equation}

\begin{equation}
\gamma=a^4(t_{c})~T^{\tau}_{\tau}(t_{c}) \label{b15}\end{equation}
and $\delta$ is an integration constant.  The homogeneity and
isotropy of the universe require that the only possibly nonzero
components of the stress tensor $T_{\mu}^{\nu}$ are
$T_{\tau}^{\tau}$ and $T_{i}^{i}$. Under this requirement and
using the trace condition (\ref{b10'}), we obtain

\begin{equation}
T_{\mu}^{\nu}= a^{-4}(H+\gamma)~diag \{1,~~ \frac{1}{3}(Q-1)~,
~~\frac{1}{3}(Q-1)~,~~ \frac{1}{3}(Q-1)\}
\label{b16}\end{equation} where

\begin{equation}
Q=\frac{T^{\mu}_{\mu}}{a^{-4}(H+\gamma)}
\label{b17}\end{equation} We can use the explicit form of the
trace anomaly (\ref{b2}) for the metric (\ref{b5}) and in terms
of the scale factor (\ref{a15}) ( see the Appendix ) to obtain
$H$ and $Q$ in asymptotic times

\begin{equation}
H(t\rightarrow
t_{0})=\frac{\lambda~a_{0}^4~n~t_{0}^{-4}}{4(n-1)}\{1-l^{4(n-1)}\}
\label{b19'}\end{equation}

\begin{equation}
Q(t\rightarrow
t_{0})=\frac{4(n-1)}{n}\{1+\frac{(3n-4)}{n}~l^{4(n-1)}\}^{-1}
\label{b19}\end{equation} We apply the same approximation used in
the two-dimensional case, namely that $l^{4(n-1)} \gg 1 $ for
$n<1$ which results in $Q(t\rightarrow t_{0}) \ll 1$. Within this
approximation and using (\ref{b15}) and (\ref{b19'}), (\ref{b16})
reduces to

\begin{equation}
T_{\mu}^{\nu}(t\rightarrow t_{0}
)=-\frac{n(3n-4)}{4(n-1)}~\lambda~l^{4n}~t_{c}^{-4}~diag \{-1,~~
\frac{1}{3}~,~~ \frac{1}{3}~,~~ \frac{1}{3}\}
\label{b20}\end{equation} This corresponds to the stress tensor
of a radiation with energy density $\rho \propto
l^{4n}~t_{c}^{-4}$. There is a specific relation between the
energy density and temperature $\rho \propto (k_{B}T)^4$ if the
radiation is thermally distributed \cite{tur}.  Therefore the
temperature of the radiation is given by $T\propto
l^{n}~(k_{B}~t_{c})^{-1}$ which is of the same order of the
result obtained in the two-dimensional case.

\section{Concluding Remarks}
We have investigated the stress tensor of a conformally invariant
quantum scalar field in a homogeneous and isotropic cosmology. We
have solved the covariant conservation equation of the stress
tensor in two different cases.  Firstly, in a two-dimensional
cosmology we have shown that the conservation equation is
consistent with existence of a thermal radiation at late times.
Secondly, in a homogeneous and isotropic universe we have been
able to generalize our results to a four-dimensional spacetime.
In this case, the general form of the stress tensor at late times
is given by the stress tensor corresponding to a thermal
radiation. In both cases the temperature of the radiation is
given in terms of a cut-off time $t_{c}$.  The cut-off is
introduced to exclude the effects of early-time cosmology, in
which a full theory of quantum gravity holds, from our
semiclassical calculations.  This result is analogous to the
result obtained in the two-dimensional schwarzschild spacetime in
the sense that in the latter case the temperature is given in
terms of a cut-off length $R_{S}$ characteristic to the
schwarzschild
solution.\\


\vspace{2cm}
{\Large{\bf Appendix }}\\\\
We consider two metric tensors $\bar{g}_{\mu\nu}$ and $g_{\mu\nu}$
which are conformally related, namely
\begin{equation}
\bar{g}_{\mu\nu}=\Omega^2 g_{\mu\nu} \label{ap1}\end{equation}
where $\Omega$ is a smooth dimensionless  spacetime function. If
$g_{\mu\nu}$ describes the Minkowski spacetime, $\eta_{\mu\nu}$,
the metric $\bar{g}_{\mu\nu}$ is
said to be conformally flat.\\
In a two-dimensional spacetime, any metric tensor can take a
conformally flat form. In this case the curvature scalar is given
by \cite{wald'}
\begin{equation} \bar{R}=-2(\Omega^{-2}\Box_{\eta}\ln \Omega)
\label{ap2}\end{equation} where $\Box_{\eta}$ is the d'Alamberian
operator in Minkowski spacetime. Substituting this into the
relation (\ref{a2}) and noting the fact that $\Omega(x)=a(\tau)$,
we obtain for the trace anomaly
\begin{equation}
T^{\mu}_{\mu}(\bar{g}_{\mu\nu})=\frac{1}{12\pi}~\frac{1}{a}~\frac{d^2
a}{dt^2} \label{ap2}\end{equation} in which we have used
(\ref{a5}) to express the derivative of the scale factor with
respect to the coordinate time $t$.  If one uses the explicit
form of the scale factor (\ref{a15}) in the relation (\ref{ap2}),
one arrives at

\begin{equation}
T^{\mu}_{\mu}(\bar{g}_{\mu\nu})=\frac{1}{12\pi}~n~(n-1)~t^{-2}
\label{ap2'}\end{equation}  In the four-dimensional case, the
trace anomaly (\ref{b2}) for the metrics $g_{\mu\nu}$ and
$\bar{g}_{\mu\nu}$ are related by \cite{brown}
$$
T^{\mu}_{\mu}(\bar{g}_{\mu\nu})=-\frac{1}{360}e^{4\omega} \{\Box
R-R_{\mu\nu}R^{\mu\nu}+R_{\mu\nu\gamma\delta}R^{\mu\nu\gamma\delta}+2R\Box\omega
+2R_{;\gamma} \omega^{;\gamma} +6\Box(\Box\omega) +8(\Box\omega
)^2
$$
\begin{equation}
-8\omega_{;\mu\nu}  \omega^{;\mu\nu} -8R_{\mu \nu}\omega^{;\mu}
\omega^{;\nu} -8\omega_{;\gamma} \omega^{;\gamma} \Box\omega
-16\omega_{;\mu\nu} \omega^{;\mu} \omega^{;\nu} \}
\label{ap3}\end{equation} where $\omega =-\ln\Omega$.  If
$g_{\mu\nu}=\eta_{\mu\nu}$, (\ref{ap3}) reduces to

\begin{equation}
T^{\mu}_{\mu}(\bar{g}_{\mu\nu})=-\frac{1}{180}e^{4\omega}
\{3\Box_{\eta}(\Box_{\eta}\omega) +4(\Box_{\eta}\omega )^2
-4\omega_{;\mu\nu} \omega^{;\mu\nu} -4\omega_{;\gamma}
\omega^{;\gamma} \Box_{\eta}\omega -8 \omega_{;\mu\nu}
\omega^{;\mu} \omega^{;\nu} \} \label{ap4}\end{equation} We may
use $\omega=-\ln a$ and (\ref{a5}) to write
$T^{\mu}_{\mu}(\bar{g}_{\mu\nu})$ in the form

\begin{equation}
T^{\mu}_{\mu}(\bar{g}_{\mu\nu})=\frac{1}{60}\{\frac{1}{a}\frac{d^{4}a}{dt^4}+
\frac{1}{a^2}\frac{d^2a}{dt^2}+3\frac{1}{a^2}\frac{da}{dt}\frac{d^3a}{dt^3}
-3\frac{1}{a^3}\frac{d^2a}{dt^2}(\frac{da}{dt})^2\}
\label{ap5}\end{equation} With the scale factor (\ref{a15}),
(\ref{ap5}) is equivalent to

\begin{equation}
T^{\mu}_{\mu}=\lambda~ t^{-4} \label{b18}\end{equation} with
\begin{equation}
\lambda=\frac{1}{30}n(n-1)(n^2-6n+3)
\end{equation}


\end{document}